\documentclass[twocolumn]{aastex631}

\usepackage{xcolor}

\newcommand*{\thethreehundred}{{\sc The Three Hundred}}

\newcommand*{\E}[1]{\ensuremath{\times 10^{#1}}}
\newcommand*{\mysub}[2]{\ensuremath{#1_{\mathrm{#2}}}}
\newcommand*{\rhocr}{\mysub{\rho}{cr}}
\newcommand*{\Msun}{\ensuremath{M_{\sun}}}

\shorttitle{A Model for Realistic Cluster X-ray Morphologies}
\shortauthors{Benyas et al.}

\graphicspath{{./}{figures/}}

\begin{document}

\title {A Generative Model for Realistic Galaxy Cluster X-ray Morphologies}

\author[0000-0001-7207-7910]{Maya Benyas}
\affiliation{Department of Physics, Stanford University, 382 Via Pueblo Mall, Stanford, CA 94305, USA}

\author[0000-0001-7718-087X]{Jordan Pfeifer}
\affiliation{Department of Physics, Brown University, Providence, RI 02912-1843, USA}

\author[0000-0002-8031-1217]{Adam B. Mantz}
\affiliation{Kavli Institute for Particle Astrophysics and Cosmology, Stanford University, 452 Lomita Mall, Stanford, CA 94305, USA}

\author[0000-0003-0667-5941]{Steven W. Allen}
\affiliation{Department of Physics, Stanford University, 382 Via Pueblo Mall, Stanford, CA 94305, USA}
\affiliation{Kavli Institute for Particle Astrophysics and Cosmology, Stanford University, 452 Lomita Mall, Stanford, CA 94305, USA}
\affiliation{SLAC National Accelerator Laboratory, 2575 Sand Hill Road, Menlo Park, CA  94025, USA}

\author[0000-0002-8800-5652]{Elise Darragh-Ford}
\affiliation{Department of Physics, Stanford University, 382 Via Pueblo Mall, Stanford, CA 94305, USA}
\affiliation{Kavli Institute for Particle Astrophysics and Cosmology, Stanford University, 452 Lomita Mall, Stanford, CA 94305, USA}

\email{mbenyas@stanford.edu,jordan\_pfeifer@brown.edu,amantz@stanford.edu}


\begin{abstract}
  The X-ray morphologies of clusters of galaxies display significant variations, reflecting their dynamical histories and the nonlinear dependence of X-ray emissivity on the density of the intracluster gas.
  Qualitative and quantitative assessments of X-ray morphology have long been considered a proxy for determining whether clusters are dynamically active or ``relaxed.''
  Conversely, the use of circularly or elliptically symmetric models for cluster emission can be complicated by the variety of complex features realized in nature, spanning scales from Mpc down to the resolution limit of current X-ray observatories.
  In this work, we use mock X-ray images from simulated clusters from \thethreehundred{} project to define a basis set of cluster image features.
  We take advantage of clusters' approximate self similarity to minimize the differences between images before encoding the remaining diversity through a distribution of high order polynomial coefficients.
  Principal component analysis then provides an orthogonal basis for this distribution, corresponding to natural perturbations from an average model.
  This representation allows novel, realistically complex X-ray cluster images to be easily generated, and we provide code to do so.
  The approach provides a simple way to generate training data for cluster image analysis algorithms, and could be straightforwardly adapted to generate clusters displaying specific types of features, or selected by physical characteristics available in the original simulations.
\end{abstract}


\section{Introduction} \label{sec:intro}

Clusters of galaxies are filled with hot plasma, the intracluster medium, which comprises most of the baryonic matter within them.
The X-ray emission from this gas, primarily from bremsstrahlung and line emission, carries a wealth of information about the dynamical histories of clusters and their internal astrophysical processes.
Since the advent of high-spatial-resolution X-ray imaging spectroscopy with the {\it Chandra} and XMM-{\it Newton} observatories, it has been possible to study in detail features such as shocks and cold fronts \citep{Markevitch0701821}, gas sloshing (e.g.\ \citealt{Ascasibar0603246, Roediger1007.4209, ZuHone1108.4427, Johnson1106.3489, Simionescu1208.2990, Paterno-Mahler1306.3520}), cavities associated with feedback from active galactic nuclei \citep{McNamara0001402, Fabian1204.4114}, and the cool, dense cores found in some clusters \citep{Peterson0512549}.
More recently, such data have been used in investigations of turbulence \citep{Churazov1110.5875, Zhuravleva1501.07271, Zhuravleva1601.02615, Zhuravleva1707.02304, Zhuravleva1906.06346, Arevalo1508.00013} and gas clumping in cluster outskirts \citep{Eckert1310.8389, Mirakhor2106.09732, de-Vries2211.07680}, the latter having been spearheaded by lower-resolution but also lower-background {\it Suzaku} observations of relatively nearby systems \citep{Simionescu1102.2429, Urban1102.2430, Urban1307.3592, Walker1205.2276, Walker1203.0486}.

The high sensitivity of X-ray emission to the dynamical processes playing out within clusters has led to features apparent in their X-ray surface brightness -- their X-ray morphology -- being used as a proxy for the underlying dynamics.
Quantitative estimates of overall symmetry \citep{Mohr1993ApJ...413..492, Buote9502002, Jeltema0501360, Nurgaliev1309.7044, Rasia1211.7040} as well as the presence of a compact, bright core \citep{Vikhlinin0611438, Santos0802.1445, Bohringer0912.4667}, or both, have been employed to identify the most dynamically ``relaxed'' clusters (e.g.\ \citealt{Mantz1502.06020}).
Other sub-populations presenting distinct morphologies that may be of special interest include major plane-of-the-sky mergers (e.g.\ \citealt{Massey1007.1924}), feedback-induced cavities, and large-scale sloshing, to name a few.

While a small number of simple estimators can be useful and effective for specific purposes, in general the rich detail and complexity of modern X-ray images of clusters far exceed what can be encoded in a few numbers.
Conversely, the presence of complex, unmodeled structure can in principle result in pathological cases where simple estimators misbehave.
This has led to the increasing adoption of ``deep learning'' methods that implicitly account for morphological characteristics while estimating quantities of interest, most often mass (e.g.\ \citealt{Ntampaka1810.07703, Yan2005.11819, de-Andres2311.02469, Ho2303.00005}).
The greater flexibility of these approaches typically comes with the requirement of significant numbers of images on which to ``train'' them, typically in the form of mock observations of clusters formed in hydrodynamical simulations.
While the variety of observed morphologies in detail poses a significant challenge to even the most advanced hydrodynamical simulations, given the wide dynamic range involved and the uncertain role of magnetic fields and feedback physics at small scales, realistic features on intermediate and large scales can be reproduced (e.g.\ \citealt{Darragh-Ford2302.10931}).
Given that such hydrodynamic simulations remain costly to produce in large numbers, there have been numerous proposed approaches to predicting the gas properties or appearance of clusters in less expensive N-body simulations, either based on simple physical models or by calibrating a translation from hydrodynamic to N-body simulations (sometimes called ``baryon pasting''; e.g.\ \citealt{Ostriker0504334, Bode0612663, Bode0905.3748, Shaw1006.1945, Sehgal0908.0540, Schneider1510.06034, Flender1511.02843, Flender1610.08029, Osato2201.02632}).

Given an existing set of hydrodynamical simulations to start from, however, the specific and limited task of generating novel X-ray images (or other observables) for training purposes can be accomplished simply and without requiring new simulations.
In this work, we employ mock X-ray images produced for clusters simulated in \thethreehundred{} project \citep{Cui1809.04622, Ansarifard1911.07878}, a suite of 324 high-resolution hydrodynamical re-simulations of halos from the dark-matter-only MultiDark simulations.
We use the known self-similarity of cluster brightness over a wide range in radius, as well as physical considerations such as isotropy, to maximize the similarity of their images before encoding their morphologies via a high-order polynomial expansion.
This representation is then diagonalized, providing a straightforward if high-dimensional basis for generating realistic cluster X-ray morphologies.
Throughout, we use the Symmetry, Peakiness and Alignment (SPA) morphological metrics of \citet{Mantz1502.06020} as a benchmark for how well this representation reproduces the information present in the original images.

The layout of the paper is as follows.
The simulated image data are described in Section~\ref{sec: data}.
In Section~\ref{sec: methods}, we review the definition of the SPA metrics that we compare to and describe the construction of the new, high-dimensional cluster image basis.
Section~\ref{sec: results} discusses the resulting distribution of morphological features as a function of dynamical state and redshift.
We consider potential applications and future directions, and conclude, in Section~\ref{sec: conclusions}.

\section{Data}
\label{sec: data}
Our data set is composed of 324 galaxy clusters from \thethreehundred{} project hydrodynamical simulations \citep{Cui1809.04622}, which are re-simulations of the most massive clusters at $z=0$ identified using the {\sc Amiga} halo finder \citep{Knollmann0904.3662} from the (1\,Gpc/h)$^3$ dark-matter-only Multi-Dark Plank 2 simulations, which assume a flat $\Lambda$CDM cosmological model with parameters $h=0.678$ and $\Omega_\mathrm{m}=0.307$ \citep{Klypin1411.4001}. These halos were re-simulated using {\sc gadget-x} \citep{Rasia1509.04247}, including baryonic and hydrodynamic physics, in smaller regions of radius $15 h^{-1}$\,Mpc. We use mock X-ray maps produced at multiple redshifts, centered on the most massive halo present in each box at a given redshift (see \citealt{Ansarifard1911.07878} for details).
In detail, these are maps of X-ray emissivity in the rest-frame energy band 0.5--2.0\,keV, integrated along the line of sight and produced along 3 orthogonal projections.
All maps are $4\times4$\,Mpc in size with a resolution of 3.9\,kpc and a line-of-sight integration length of 10\,Mpc, regardless of redshift.
The maps were converted to images of X-ray surface brightness in the observer-frame 0.6--2.0\,keV band using temperature-dependent K-corrections (see \citealt{Darragh-Ford2302.10931}).
We employ such maps generated at redshifts 0.333, 0.592 and 0.986, between which the mass range of the cluster sample evolves from $M_{200}\sim (0.6$--$12)\E{14}\,\Msun$ to $\sim (2.5$--$25)\E{14}\,\Msun$.

\section{Methods}
\label{sec: methods}

In this section, we detail the development of our new generative model of cluster X-ray morphology.
To provide a baseline for how well the model encodes key morphological information, we first extract the SPA metrics for each cluster and projection, as described in Section~\ref{sec:spa}.
Sections~\ref{sec:cximb}--\ref{sec:pca} then step through the procedure used to define our model.

Below, we employ the conventional ``overdensity'', $\Delta$, which jointly defines a family of characteristic cluster masses and radii as
\begin{equation} \label{eq:overdensity}
    M_\Delta = \frac{4\pi}{3} \Delta \rhocr(z) r_\Delta^3,
\end{equation}
where $\rhocr(z)$ is the critical density at the cluster's redshift.
Broadly speaking, archival data from modern X-ray observatories like {\it Chandra} and XMM-{\it Newton} can probe the 2-dimensional shape of many clusters out to approximately $r_{2500}$ and firmly detect emission out to $r_{500}\approx 2.3\,r_{2500}$, although there is naturally variation with exposure time, extent, redshift and overall luminosity.
The masses of the simulated clusters noted in Section~\ref{sec: data} correspond to values of $r_{500}$ evolving from $\sim 0.8$--1.3\,Mpc to $\sim 0.5$--0.8\,Mpc between redshifts 0.333 and 0.986.

\subsection{SPA Metrics}
\label{sec:spa}

\citet{Darragh-Ford2302.10931} describe the procedure for computing SPA metrics of the mock X-ray images from \thethreehundred{} simulations.
We reproduce their methods, with the exception that the metrics are computed only within a circular region of radius $r_{500}$ (see Section~\ref{sec:standardization}) rather than on the complete images.
This allows a fair comparison of the metrics derived from the original images and from images produced from the model, which is only defined within a circle.
Below, we review the most important aspects of SPA for this discussion; the algorithm is described in full detail by \citet{Mantz1502.06020}.

A key aspect of the SPA approach is to approximately remove the redshift and temperature (i.e.\ mass) dependent scaling of the observed surface brightness of a cluster's X-ray emission (\citealt{Mantz1502.06020}. As modified by \citealt{Darragh-Ford2302.10931}), this scaling factor is
\begin{equation} \label{eq:sbscal}
  f_S = K(z,T) \frac{E(z)^3}{(1+z)^4} \left(\frac{kT}{\mathrm{keV}}\right) \left(\frac{5.7\E{-8}\,\mathrm{erg}}{\mathrm{Ms}\,\mathrm{cm}^2\,(0.984'')^2}\right).
\end{equation}
Here, $K$ converts between bolometric flux in energy units and photon flux in the observed energy band,\footnote{When applied to real data, this term would also account for Galactic absorption.} $kT$ is the temperature of the cluster gas at approximately $r_{2500}$, and $E(z)=H(z)/H_0$ is the normalized Hubble parameter at the cluster's redshift.
In units of $f_S$, the regions of similar observed surface brightness, $S$, in different clusters correspond to broadly comparable overdensities, with the exception of central regions where the presence or absence of a cool core causes significant variations in brightness.

The first step of the SPA algorithm is to define a global cluster center as the median photon position of the flat-fielded, background-subtracted image. Given this center, the SPA metrics themselves can be understood as:
\begin{enumerate}
    \item Peakiness ($p$) quantifies the average scaled surface brightness in the cluster center, defined as radii where an azimuthally averaged surface brightness profile exceeds $\sim0.05 f_S$.
    \item Symmetry ($s$) quantifies the agreement between the global center and the centers of 5 ellipses, fit to isophotes that logarithmically span the range 0.002--0.05\,$f_S$.
    \item Alignment ($a$) quantifies the agreement among the centers of the above ellipses.
\end{enumerate}
A cluster image is categorized as relaxed if it simultaneously exceeds specific thresholds in each metric ($s>0.87$, $p>-0.82$, $a>1.00$); complete details and formulae are given by \citet{Mantz1502.06020}.

\subsection{Constructing an Image Basis}
\label{sec:cximb}

Our starting point is the set of Zernike polynomials (ZPs), comprising an orthogonal basis of functions over the unit circle that can, in principle, encode an arbitrary circumscribed image.
\citet{Capalbo2009.04565} have used this basis to quantify the morphology of simulated maps of the thermal Sunyaev-Zel'dovich effect of \thethreehundred{} clusters, demonstrating a relationship between the amplitude of the asymmetric polynomial coefficients and simulation-based metrics for dynamical state.
From the standpoint of developing a generative model, the main challenge is to identify and parameterize the subspace of this general model space containing realistic cluster morphologies, as defined by the simulated images.
This, along with computational limitations, necessarily requires us to choose a maximum order of polynomial (corresponding to a smallest spatial scale) to include in the modeling.
Given the large dynamic range of surface brightness (typically 4--5 orders of magnitude within $r_{500}$) and morphologically relevant spatial scales ($\sim10$--1000\,kpc) in X-ray images of clusters, these are non-trivial decisions.

\begin{figure*}
    \centering
    \includegraphics[width = .9  \textwidth]{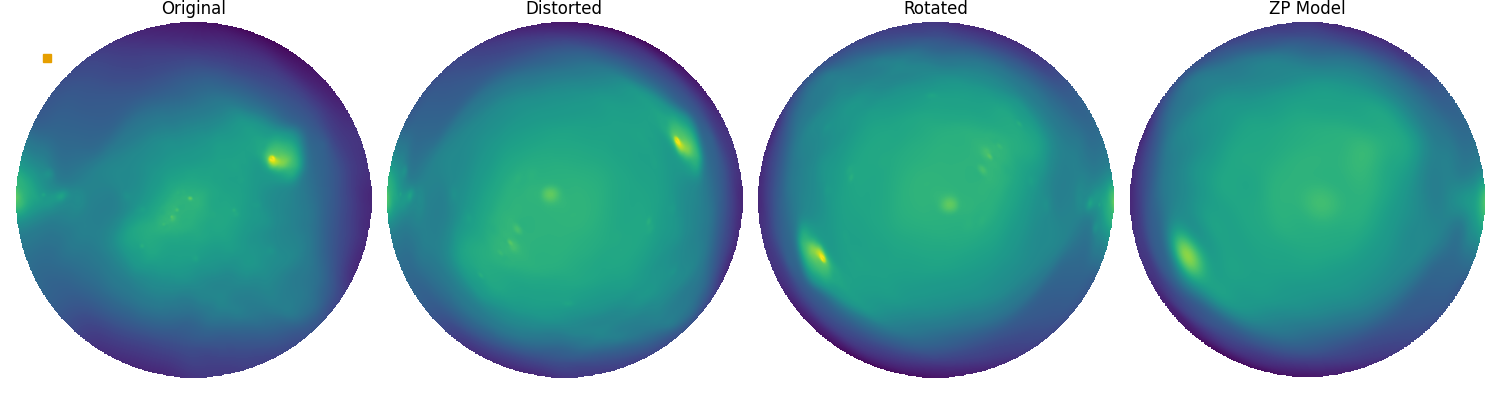}
    \caption{Our analysis process performed on example image (identified by an orange square in later figures). Note that all images, including the ``original'', are log-scaled for clarity. Images are first standardized as described in Section \ref{sec:standardization}, and then radially distorted such that the center is magnified (Section \ref{sec:radial}). All images in our dataset are then rotated by an angle selected as described in Section~\ref{sec:zpfit}. The rightmost panel shows the image corresponding to the ZP representation of this processed input.
    }
    \label{fig:flowchart}
\end{figure*}

The above considerations provide key context for our overall strategy in developing a morphological image basis, which is detailed in the remainder of this section.
 First, we apply a standardization to the original images, with the aim of leveraging physical considerations to minimize the image-to-image variation \emph{before} encoding anything in the polynomial basis (Section~\ref{sec:standardization}).
To reduce the polynomial order required to adequately encode the presence of sharply peaked cool cores, we then apply a non-linear radial distortion (Section~\ref{sec:radial}).
In Section~\ref{sec:pca}, we fit the standardized, distorted images using the ZP basis, and apply Principal Component Analysis (PCA) to identify a lower-dimensional model subspace encoding the most important image-to-image variations.
Figure~\ref{fig:flowchart} illustrates the procedure.

\subsubsection{Image Standardization}
\label{sec:standardization}

The more we can reduce the variation among images a priori on the basis of physical considerations, the lower-dimensional our final morphology model need be.
To take a simple example, images that differ only by a relative rotation should ideally be encoded by an identical set of model parameters apart from one.
Since this behavior does not naturally emerge from the procedure described in the previous section, we explicitly apply a series of simple, standardizing transformations to the original images at the outset.
That is, our approach is opposite to that normally used in computer vision: rather than augmenting the training data with many transformed replicates (e.g.\ rotated versions of each image), we instead define a preferred rotation with respect to which the model is defined.
The final set of model parameters then need only be supplemented with a position angle.
A similar approach can be taken with other simple image transformations, namely centering (translation), extent (zoom) and scaling.
However, while it is desirable for the part of the model encoding cluster morphology to be invariant under rotation, these other transformations are physically significant, providing us the opportunity to build physical considerations into the model at a fundamental level.
\begin{enumerate}
\item Scaling: Exactly the same considerations that motivate the scaling of surface brightness by the self-similar factor $f_S$ in the SPA approach apply here (Section~\ref{sec:spa}). Once divided by $f_S$, surface brightness profiles of clusters are observed to have relatively small scatter (few tens of per cent; \citealt{Mantz1502.06020}), apart from small radii where cool cores can boost the brightness by an order of magnitude. We also logarithmically scale the images, reducing the dynamic range of the data to be modeled from several orders of magnitude to, typically, a single order of magnitude, making them much more simply encoded by a basis of polynomial functions.
\item Centering: The centers of the simulated images are, by construction, the position of the gravitational potential minimum. We maintain this centering for the circular region within which the modeling takes place.
\item Radial extent: We define $r_{500}$, as computed from the spherically averaged enclosed mass profile about the above center, as the radial extent of the model.
Subsequent analysis is thus limited to this circular region of the original image. As noted above, this choice of radial extent should provide a model that comfortably applies to the typically visible portions of clusters observed by {\it Chandra} and XMM-{\it Newton}.
\item Rotation: Following the logic above, we define a preferred rotation; however, in practice, this is applied after the radial distortion introduced in the next section. We therefore defer further discussion to Section~\ref{sec:zpfit}.
\end{enumerate}

\subsection{Radial Distortion}
\label{sec:radial}

The smallest scale represented by the Zernike basis shrinks only modestly quickly with order, while the computational expense increases super-linearly.
The sharpness of the brightness peaks of cool cores, which in extreme cases have power-law like profiles down to the resolution limit of current X-ray observatories, thus poses a challenge, even with logarithmic scaling of the images.
To make the representation of small-scale features in cluster cores more tractable, we therefore introduce a non-linear radial distortion of the standardized images.
Specifically, the transformation is
\begin{equation} \label{eq:radial}
x \rightarrow x' = A \ln\left(1 + \frac{x}{\xi}\right),
\end{equation} 
where $x=r/r_{500}$, $\xi=0.05$, and $A$ normalizes the new radial coordinate such that $x'=1$ corresponds to $x=1$.
This has the effect of greatly magnifying the cluster center, at the expense of the outskirts, with the transformation approaching linearity for $x \gg \xi$.
The distortion thus does not help to resolve comparably small features outside of the cluster center, but these are relatively rare in real clusters (e.g., the merging cluster Abell~115).
Figure~\ref{fig:flowchart} shows the impact of this procedure on an example image.
(Note that we reverse this distortion before displaying model images in all other figures.)

\subsection{Zernike Polynomial Fitting and Rotation}
\label{sec:zpfit}

Once a simulated image is standardized and distorted, we use the {\sc zernike} Python package to determine the ZP coefficients up to a maximum order of $n=30$ (a total of 496 coefficients)
by numerically evaluating the inner product of each polynomial with the image.

Once the ZP coefficients for an image have been found, we standardize its rotation.
This step is done after finding the coefficients for convenience, as the transformation of ZP coefficients under rotation is particularly simple.
To define a preferred rotation, we first find the position angle that maximizes the $Z_2^2$ coefficient, and then break the remaining symmetry (between two angles separated by $180^\circ$) by maximizing the $Z_1^1$ coefficient.
This is notionally similar, though not identical, to aligning the major axis of an elliptical cluster with the $x$ axis, with more brightness (e.g.\ due to substructure) lying in the positive $x$ direction than the negative $x$ direction.

\subsection{Principal Component Analysis}
\label{sec:pca}

The final step of the model development is to perform PCA on the ZP coefficients of the standardized, distorted and rotated images.
The motivation here is to identify a new, lower-dimensional model that spans the space of realistic cluster morphologies (as described by the simulated images) without the additional complexity required to generate an arbitrary image.

Before finding the principal components, we augment the data using the one simple image transformation not addressed in the preceding sections: reflection of the standardized images over the $x$ axis.
We follow the standard practice of centering the input data (subtracting the mean and rescaling to unit variance), and use the {\sc scikit-learn} Python package to carry out the PCA.

\section{Results and Discussion}
\label{sec: results}

\begin{figure*}
  \centering
  \includegraphics[width=0.95\textwidth]{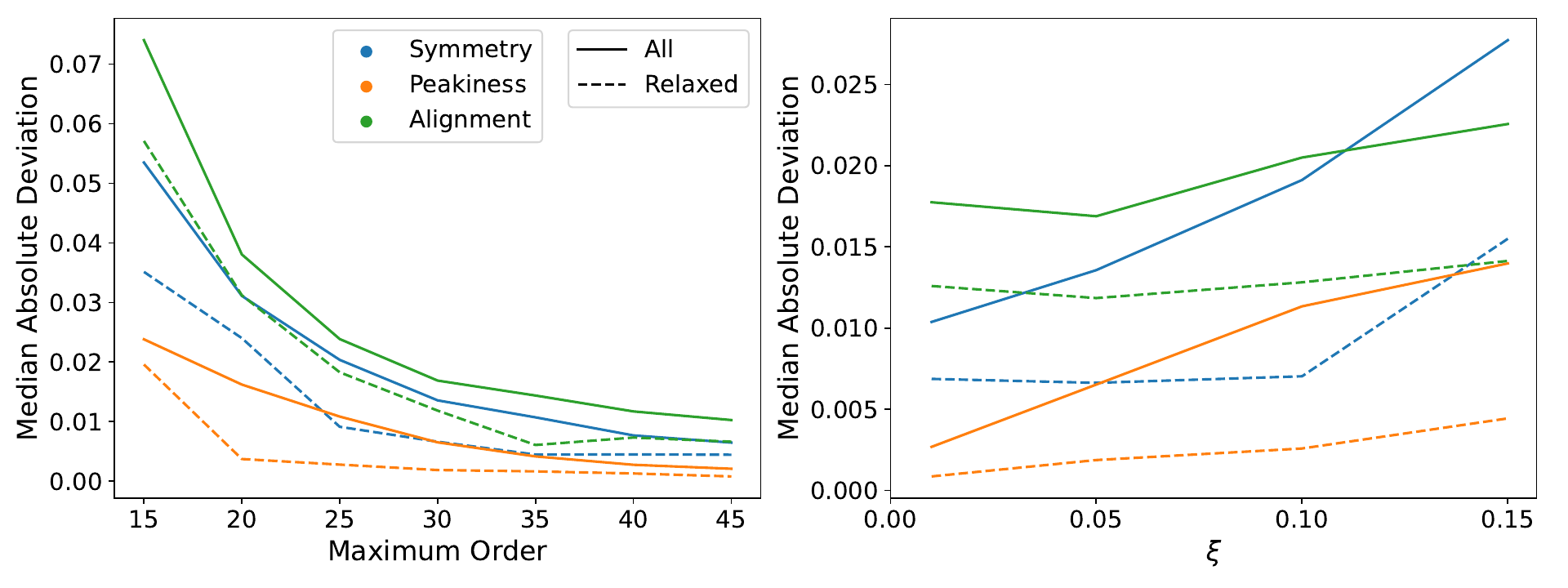}
  \caption{
    MADs of the SPA metrics (comparing original and modeled images) are plotted against the maximum ZP order and magnification parameter, the default values of which are $n=30$ and $\xi=0.05$, respectively.
    Solid and dashed curves respectively show the MADs for all simulated images in the $z=0.333$ snapshot, and only relaxed images.
  }
  \label{fig: MAD of spa w/order & xi}
\end{figure*}

\begin{figure}
  \centering
  \includegraphics[width=0.45\textwidth]{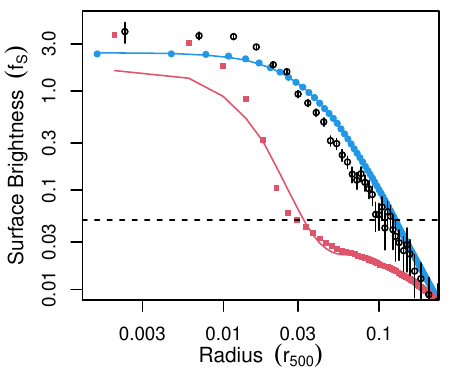}
  \caption{
    Scaled surface brightness profiles of the cores of SPT~J2344, the peakiest cluster observed by \citep[][open black circles]{Mantz1502.06020}, and two similarly peaky simulated clusters (filled blue circles and red squares; all have $p>-0.5$).
    The latter, the peakiest simulated cluster, has an extreme value of the first principal component, PC0, of $\sim11$.
    Solid curves show profiles of the images reconstructed from the ZP model.
    The dashed line shows the threshold relevant for the definition of peakiness \citet{Mantz1502.06020}.
  }
  \label{fig: peakiest}
\end{figure}

\subsection{Impact of the Maximum Order and Magnification}
\label{sec:tuning}

As described in Section \ref{sec: methods} above, the procedures of ZP fitting and radial distortion require us to choose a maximum order of ZP and a magnification parameter, respectively. Both cases demand a compromise; in choosing a maximum ZP order, we strike a balance between computational overheads and the ability to reproduce small-scale features. Varying the magnification parameter, we trade detail at the outskirts of the cluster for additional resolution in cluster cores.
To explore these trade-offs, we compare the original and modeled images of the $z=0.333$ snapshots, using the SPA metrics as proxies for how well important morphological features are encoded by the model.

Figure~\ref{fig: MAD of spa w/order & xi} shows how the median absolute deviation (MAD) of each metric varies when either the maximum ZP order or the magnification parameter is changed.
Here deviation refers to the differences between metrics computed from the original images and images reconstructed from the ZP model.
As expected, all MADs decrease monotonically with increasing order.
As central magnification increases ($\xi$ decreases), we also see greater agreement, although gains within the explored range are relatively modest.
The model is quantitatively better at reproducing the SPA metrics for relaxed clusters, which is intuitive given their overall simpler morphologies.
In this context, we note that typical statistical uncertainties on SPA measurements for real clusters with {\it Chandra} are $\sim0.05$ in peakiness and $\sim0.10$ in symmetry and alignment \citep{Mantz1502.06020}.

Despite this agreement, we note that there are features that appear in some simulated clusters that are not seen in real data, in particular in cluster cores (see also \citealt{Darragh-Ford2302.10931}).
Figure~\ref{fig: peakiest} compares the surface brightness profile of the peakiest simulated cluster, which displays a relatively large, flat core on which an exceptionally sharp peak is superimposed, with the much smoother profile of the peakiest real cluster from \citet{Mantz1502.06020}.
The extremely compact (of order 10\,kpc) cores in some simulated clusters, which are an expected consequence of their limited resolution \citep{Rasia1509.04247}, are not well reproduced by the model with any combination of maximum order and magnification that we considered.
On the other hand, the simulations also include peaky systems that better resemble real clusters, as shown in the figure, and that are easily described by the baseline model with $n=30$ and $\xi=0.05$.

\subsection{The PC Representation}

\begin{figure*}
  \centering
  \includegraphics[scale=0.45]{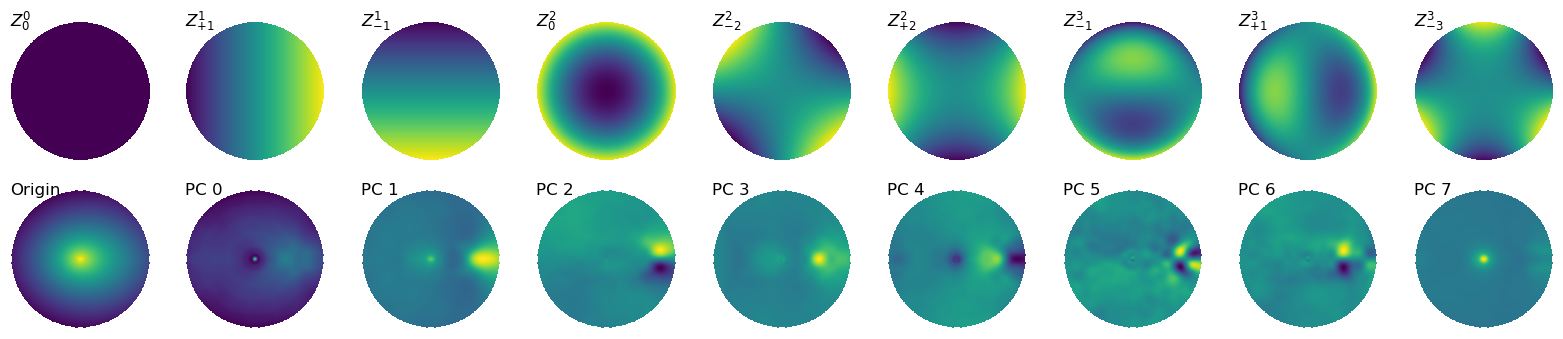}
  \vspace{-2ex}
  \caption{%
    Top row: visualizations of the first nine ZPs (in canonical order).
    Bottom row: the leftmost image corresponds to the origin of PC space.
    The others are difference images between unit vectors along the first 8 PCs and the origin.
    Color tables are independent in each panel and are symmetric about zero (except for the origin image).
  }
  \label{fig:archetypes}
\end{figure*}

The bottom row of Figure~\ref{fig:archetypes} visualizes the most important features encoded by the PCA.
The first of these images corresponds to the origin of PC space; that is, the average cluster image after standardization.
Each of the remaining images is the difference (in log-brightness) between a model image corresponding to a unit vector along a given PC and the origin.
For comparison, the top row shows the first nine ZPs (the origin of ZP space corresponds to an image of all zeros, visually identical to the constant $Z_0^0$ polynomial).
While both representations ultimately contain the same information, the origin and basis directions of the PC representation, by construction, more closely resemble visually identifiable features of the input images.

The impact of the standard rotation imposed in Section~\ref{sec:zpfit} is evident here, with many of the features encoded in these first PCs appearing to be concentrated near the positive $x$ axis.
The shape of the cluster core is the other clear feature represented in these first PCs, with several of them modulating the central brightness on different scales; PC0 in particular appears related to the unphysically sharp peaks atop flatter cores noted in Section~\ref{sec:tuning} (see also Figure~\ref{fig: peakiest}).

\begin{figure*}
    \centering
    \includegraphics[scale=0.365]{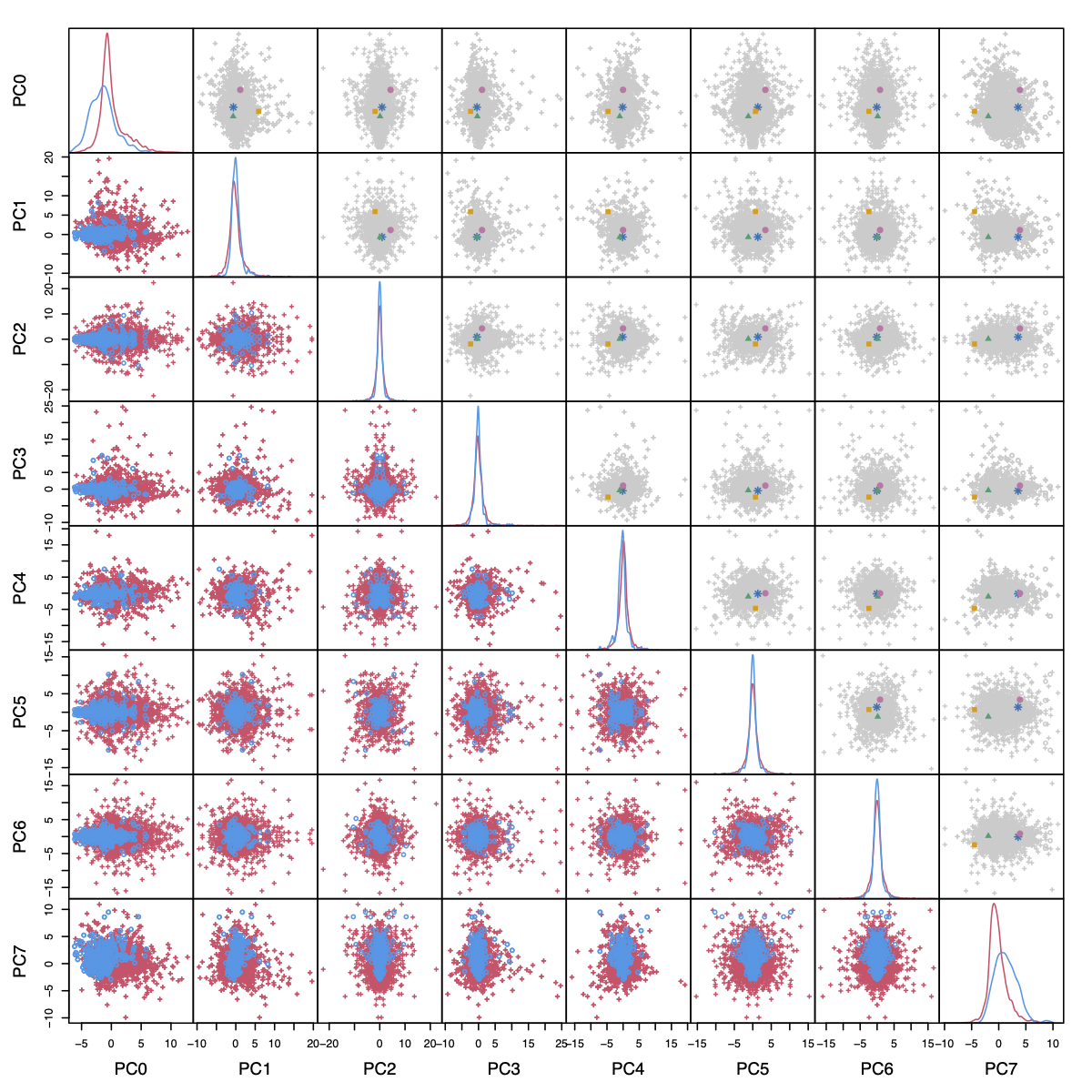}
    \vspace{-2ex}
    \caption{%
    Coefficient distributions of the first eight PCs for the simulated cluster images (including their reflections).
    Diagonal: smoothed histograms for each PC, distinguishing images classified as relaxed (blue) and unrelaxed (red) by the SPA metrics.
    Lower-left triangle: pairwise distributions of coefficients for relaxed (blue circles) and unrelaxed (red crosses) images.
    Upper-right triangle: as above, but colors and symbols distinguish 4 example images to be referred to later in the text, with all other points shown in gray. See Figure~\ref{fig:reconstruct_exemplars} for images of the exemplars.
    }
    \label{fig:PC vs PC}
\end{figure*}

Histograms and pairwise distributions of coefficients for these PCs are shown in Figure~\ref{fig:PC vs PC}, distinguishing between relaxed (blue) and unrelaxed (red) images.
Note that, by construction, the average of each coefficient is zero;
the correlations among coefficients are also effectively zero, although this need not have been the case (correlation coefficients $<10^{-14}$ in absolute value).
In most cases, the distributions for both relaxed and unrelaxed clusters peak at zero, with the main difference being a larger scatter for unrelaxed clusters.
However, within this abbreviated group of PCs, we see clear differences in PC0 (encoding an unphysical peak, as noted above, as well as an excess at relatively small radii; here relaxed clusters prefer \emph{negative} values) and PC7 (encoding a larger, bright peak; here relaxed clusters prefer \emph{positive} values).
Due to the augmentation of the data by reflecting over the $x$ axis in Section~\ref{sec:pca}, each PC is either perfectly symmetric (e.g.\ 0, 1, 3, 4, 7) or anti-symmetric (e.g.\ 2, 5, 6) with respect to $x$-axis reflection.
In the upper-right triangle of the figure, different colors/symbols show the PC coefficient values for 4 example images chosen for their differing visual appearances (see Figure~\ref{fig:reconstruct_exemplars}).

\begin{figure*}
    \includegraphics[width=\linewidth]{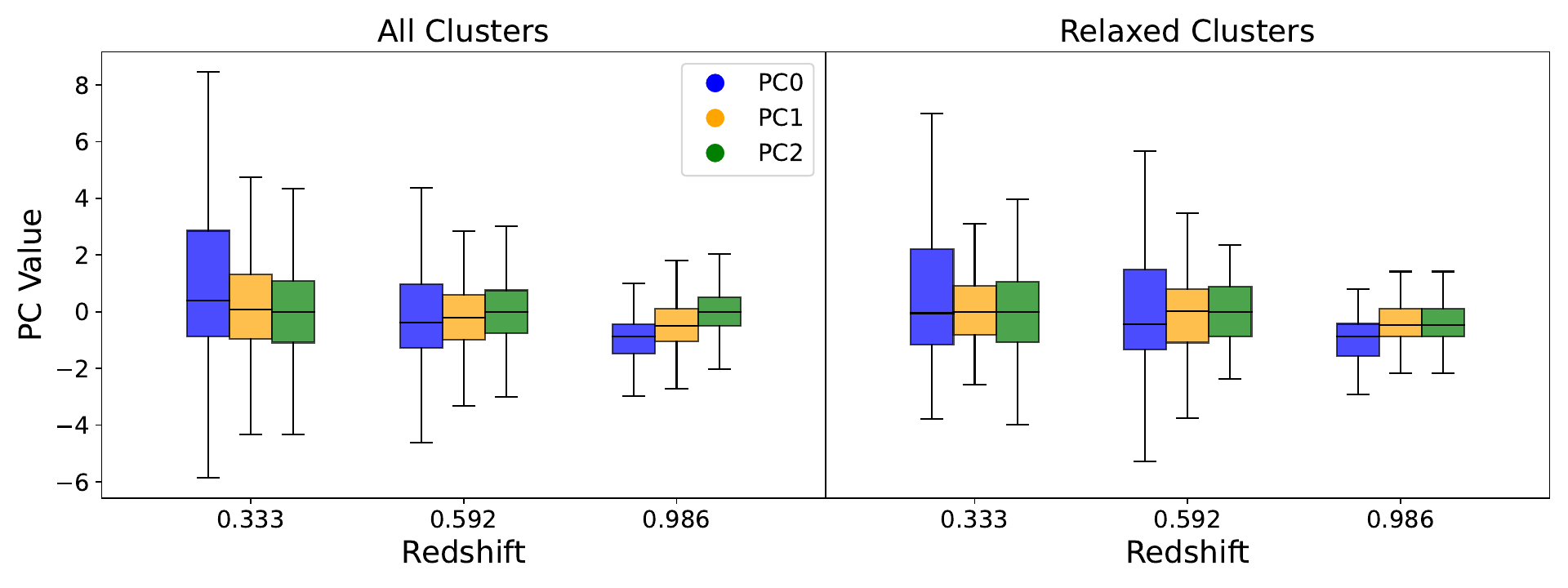}
    \caption{ Box and whisker plots depicting the redshift dependence of the first three principal components. PC0 has consistent decreases in median value with redshift, PC1 has less pronounced changes, and PC2 has consistent median values across redshifts. All three PCs shown have decreased range with increased redshift.
    }
    \label{fig:redshift}
\end{figure*}

Beyond this overall distribution, we observe formally significant trends as a function of redshift in the $x$-reflection symmetric PCs\footnote{The anti-symmetric PCs necessarily have a distribution that is perfectly symmetric about zero at all redshifts.} (Figure~\ref{fig:redshift}) when fitting a linear model, although changes in the mean of each PC are generally small compared with the scatter.
The scatter in all coefficients consistently decreases with increasing redshift.
This seems to imply that clusters look more similar to one another at high redshifts, regardless of their specific morphologies.
It is possible that the overall trends are related to growth, given that $r_{500}$ is larger at smaller redshifts; we might therefore expect that some features -- those that do not scale with the cluster size -- show up in different PCs at different redshifts, even if they are present at the same level.
Similarly, the contrast between cool cores and the self-similar cluster profile on larger scales is known to decrease with redshift \citep{Mantz1502.06020, McDonald1702.05094}.
The decreasing scatter might be explained by the natural reduction in sound crossing time at correspondingly higher redshifts, although the higher merger rate at earlier times might be expected to counteract this to some extent.

\subsection{Fidelity as a Function of the Number of PCs}

\begin{figure*}
  \centering
  \gridline{%
    \includegraphics[width=0.45\textwidth]{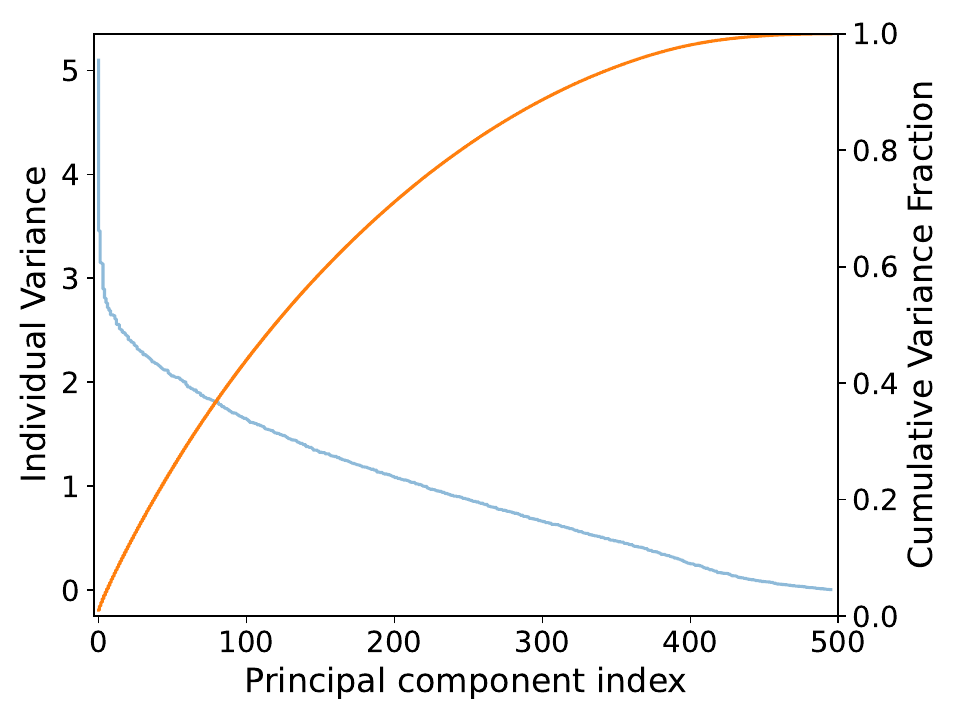}
    \includegraphics[width=0.45\textwidth]{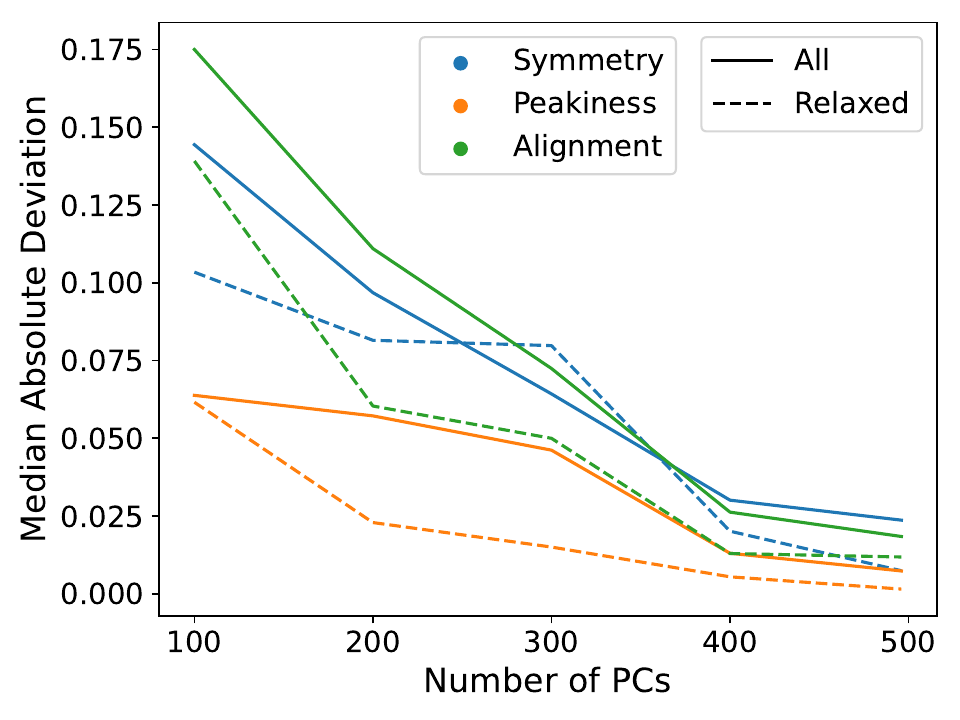}
  }
  \caption{
    Left: Individual (blue line) and cumulative (red line) variance explained by each principal component of the model.
    Right: Similarly to Figure~\ref{fig: MAD of spa w/order & xi}, MADs of the SPA parameters (comparing original and modeled images), as a function of the number of PCs used in the model image reconstruction.
    A single projection of each $z=0.333$ simulated cluster is used in this case.
  }
  \label{fig:Npcs}
\end{figure*}

We next consider how much dimensional reduction the PC representation might provide, with the caveat that this will depend on the particular application.
The left panel of Figure~\ref{fig:Npcs} shows a standard PCA diagnostic, the variance explained by each PC (individually as well as cumulatively).
After the first 4 PCs, the explained variance decreases slowly, with e.g.\ almost 300 PCs required to explain 90 per cent of the total variance.

\begin{figure*}
    \centering
    \includegraphics[scale=0.45]{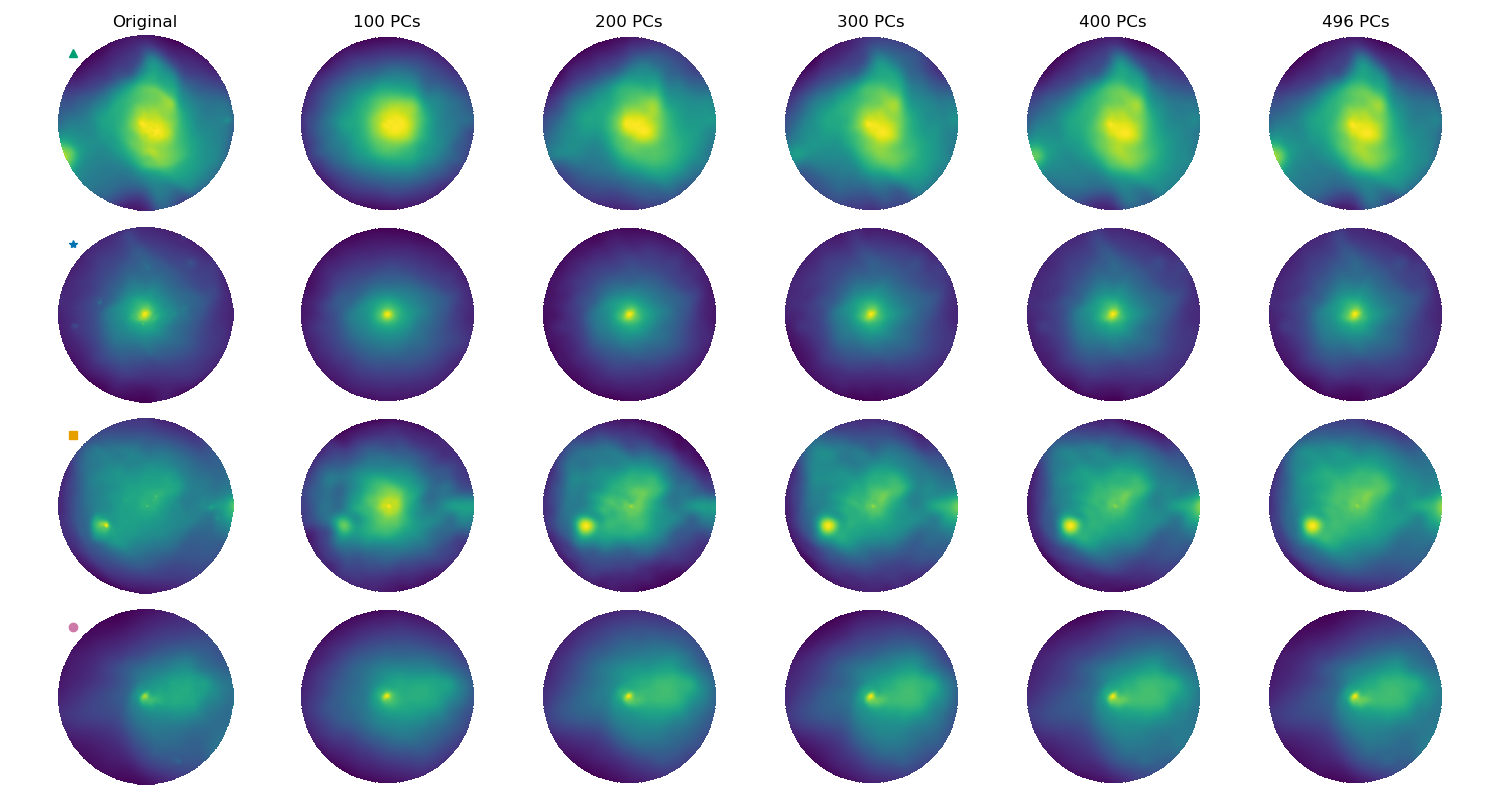}
    \caption{Shown in the left column are four example clusters, followed by partial reconstructions of each using various numbers of PCs (in order). Each image is color-scaled separately.}
    \label{fig:reconstruct_exemplars}
\end{figure*}

The slowness of this ``convergence'' can be seen in Figure~\ref{fig:reconstruct_exemplars}, which shows 4 example images (identified in Figure~\ref{fig:PC vs PC}) as modeled by increasing numbers of PCs, in order.
While the broad shape of each image is established with relatively few PCs, and resemblance to the original images clearly improves with additional PCs, these improvements are visually incremental after 200--300 PCs.
More quantitatively, we can compare the SPA metrics obtained from the original images with those computed from model images using a subset of PCs, as in the right panel of Figure~\ref{fig:Npcs}, which similarly shows steady improvement up until 300--400 PCs are in use.
As previously, we note that measurement uncertainties in these metrics from real data are typically $\sim 0.05$--0.10.

\section{Conclusions}
\label{sec: conclusions}

In this work, we have explored one strategy for defining a basis of cluster X-ray morphologies, based on simulated X-ray emissivity maps from \thethreehundred{} project.
Our approach aims to maximize the similarity of the simulated images at the outset by taking advantage of physical insights such as the cluster mass and center, the approximately self-similar scaling of surface brightness, and rotational alignment with surrounding large-scale structure (as manifest in overall ellipticity and dipolar asymmetry).
After this standardization, images are distorted to reduce the sharpness of features in cluster centers, and fitted by a high-order set of polynomial functions.
PCA of the polynomial coefficients provides a basis set of images that are more naturally interpretable in the context of cluster images than the original polynomials, even though individual PCs generally do not correspond to identifiable, physically meaningful features.

\begin{figure}
    \centering
    \includegraphics[width = .45  \textwidth]{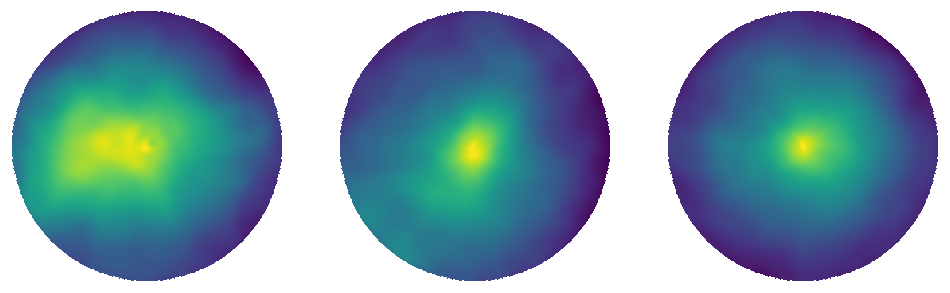}
    \caption{Three fictional cluster images generated by choosing Principal Component coefficients at random from a Gaussian distribution for each PC, as well as a random rotation angle.}
    \label{fig:fictional}
\end{figure}

One immediate application of this approach is the generation of novel, realistic cluster images.
Figure~\ref{fig:fictional} shows three examples, produced by randomly drawing PC coefficients from normal distributions with mean zero and standard deviations equal to the scatter in the simulated data.\footnote{Code used to do so can be found at \url{https://github.com/kipac-xoc/cximb}.}
One could compute a more detailed description of the high-dimensional PC coefficient distribution, but even this simple method produces complex images that qualitatively resemble the simulated clusters.
Such mock images may provide a useful alternative to simple, parametric models for testing for cluster morphology or feature detection algorithms (e.g.\ cavity-finding; \citealt{Fort1712.00523, Plsek2304.05457}).
Our results thus provide a near-instantaneous alternative to running new hydrodynamical simulations, as well as ``baryon pasting'' approaches (which require N-body simulations) when \emph{only} novel X-ray images, as opposed to complete physical descriptions of clusters, are needed.
Naturally, care must be taken, as unphysical features present in the original simulations will also be reproduced;
on the other hand, it may be straightforward to suppress such features if they are primarily represented by a single PC (e.g.\ PC0).
With appropriate restriction of the input data, one could straightforwardly construct more specific models, e.g.\ for example to generate images of only relaxed clusters, only first-pass mergers, etc.
Application to other wavelengths would also be straightforward given input simulations, though questions such as the appropriate maximum polynomial order and radial distortion would need to be revisited for those cases.

Given the high dimensionality of the model, its use with real X-ray data will likely require application-specific dimensionality reduction.
For example, we find that more than half of the almost 500 PCs must be included for the SPA morphology metrics to be faithfully reproduced (within typical measurement uncertainties) in a model image.
This is not entirely surprising, given that the PCA is in no way informed by SPA, nor more generally by knowledge of what features distinguish ``relaxed'' and ``unrelaxed'' morphologies.
As the bulk of the observational literature relies on as few as 1--3 quantitative metrics of X-ray morphology (e.g.\ \citealt{Postman1106.3328, Mahdavi1210.3689, Martino1406.6831, Mantz1502.06020, Campitiello2205.11326}), it seems reasonable that a small number of informed combinations of PCs (or some other decomposition of the original parameter space) may be sufficient for many purposes.
Future work will investigate methods for producing such low-dimensional models, tailored to identifying specific morphological types from real images.

\begin{acknowledgments}
MB was supported by Stanford University's Physics Research Program.
JP was supported by the U.S. Department of Energy, Office of Science, Office of Workforce Development for Teachers and Scientists (WDTS) under the Science Undergraduate Laboratory Internships Program (SULI).
We acknowledge support from the U.S. Department of Energy under contract number DE-AC02-76SF00515 to SLAC National Accelerator Laboratory.
This work has been made possible by the ``The Three Hundred'' collaboration (\url{https://www.the300-project.org}).
\end{acknowledgments}

\smallskip
\software{%
  scikit-learn \citep{scikit-learn},
  zernike \citep{AntonelloJOSAA.32.001160}
}

\bibliographystyle{aasjournal}
\newcommand*{\science}{Science}


\end{document}